\documentclass[%
 reprint,
superscriptaddress,
 amsmath,amssymb,
 aps,
 showkeys
]{revtex4-2}
\usepackage{gensymb}
\usepackage{booktabs}
\usepackage{graphicx}
\graphicspath{{./Figures/}}
\usepackage{dcolumn}
\usepackage{multirow}
\usepackage{bm}
\usepackage{times}
\usepackage{lipsum}

\usepackage[colorlinks = true,
            linkcolor = blue,
            urlcolor  = blue,
            citecolor = blue,
            anchorcolor = blue]{hyperref}

\begin{document}

\preprint{APS/123-QED}

\title{
Active Learning for Conditional Inverse Design with Crystal Generation \\and Foundation Atomic Models}

\author{Zhuoyuan Li}
\author{Siyu Liu}
\author{Beilin Ye}
\author{David J. Srolovitz}\email{srol@hku.hk}
\author{Tongqi Wen}\email{tongqwen@hku.hk}
\affiliation{Center for Structural Materials, Department of Mechanical Engineering, The University of Hong Kong, Hong Kong SAR, China}
\affiliation{Materials Innovation Institute for Life Sciences and Energy (MILES), HKU-SIRI, Shenzhen, China}

\date{\today}

\begin{abstract}

Artificial intelligence (AI) is transforming materials science, enabling both theoretical advancements and accelerated materials discovery. Recent progress in crystal generation models, which design crystal structures for targeted properties, and foundation atomic models (FAMs), which capture interatomic interactions across the periodic table, has significantly improved inverse materials design. However, an efficient integration of these two approaches remains an open challenge. Here, we present an active learning framework that combines crystal generation models and foundation atomic models to enhance the accuracy and efficiency of inverse design. As a case study, we employ Con-CDVAE to generate candidate crystal structures and MACE-MP-0 FAM as one of the high-throughput screeners for bulk modulus evaluation. Through iterative active learning, we demonstrate that Con-CDVAE progressively improves its accuracy in generating crystals with target properties, highlighting the effectiveness of a property-driven fine-tuning process. Our framework is general to accommodate different crystal generation and foundation atomic models, and establishes a scalable approach for AI-driven materials discovery. By bridging generative modeling with atomic-scale simulations, this work paves the way for more accurate and efficient inverse materials design.

\end{abstract}

\maketitle

\section{Introduction}
\label{sec:intro}

Crystalline materials, characterized by their periodic atomic arrangements, exhibit distinct physical and chemical properties for a wide range of applications in structural materials, electronics, optics, and catalysis. The discovery of novel crystalline materials is fundamental to advancing materials science, enabling further property optimization through composition tuning, strain engineering, and microstructure design~\cite{raabe2023accelerating}. However, conventional crystal structure search and discovery methods~\cite{oganov2011evolutionary,wang2010crystal}, which rely heavily on experimental synthesis and computational simulations, are often time-consuming and resource-intensive, limiting the speed of innovation~\cite{pyzer2022accelerating, cgcnn_xie2018crystal}. The emergence of artificial intelligence (AI) is reshaping the paradigm of materials discovery by leveraging large-scale datasets from experiments and theoretical studies. Machine learning (ML) and deep learning approaches have demonstrated remarkable success in accelerating crystal structure prediction, optimizing material properties, and guiding experimental synthesis~\cite{bartok2017machine, himanen2019data, chanussot2021open}. These data-driven techniques offer a more efficient and cost-effective alternative to traditional methods, enabling automated property predictions and inverse design strategies that streamline the discovery of novel crystalline materials.

In AI-driven crystal structure discovery, research is primarily divided into forward and inverse problems. The forward problem involves predicting material properties from given crystal structures and establishing structure-property relationships, while the inverse problem focuses on generating crystal structures with desired properties. 
Forward modeling approaches can be broadly categorized into direct and indirect methods. 
Direct methods employ graph neural networks (GNNs), such as SchNet~\cite{schnet_schutt2017schnet}, CGCNN~\cite{cgcnn_xie2018crystal}, and MEGNet~\cite{megnet_chen2019graph}, to map crystal structures directly to material properties using extensive datasets. 
These models learn complex structure-property correlations and enable rapid property predictions. 
In contrast, indirect methods leverage machine learning force fields (MLFFs) to approximate potential energy surfaces based on high-quality first-principles calculations~\cite{wen2022deep}. 
MLFFs enable atomistic simulations that capture both thermodynamic and kinetic behaviors, providing deeper insights into material properties beyond static descriptors~\cite{PhysRevMaterials.6.113603}. 
Recent advances in foundation atomic models (FAMs) further enhance AI-driven materials design by developing foundation models that describe atomic interactions across most elements in the periodic table~\cite{batatia2024foundationmodelatomisticmaterials,zhang2024dpa,chen2022universal,deng2023chgnet}. These models streamline property predictions for multi-component materials, enabling more efficient exploration of complex chemical spaces. Both direct and indirect approaches, particularly when integrated with FAMs, play a crucial role in accelerating property screening and guiding the discovery of novel materials~\cite{wang2025crystallinematerialdiscoveryera}.

In contrast to property prediction, crystal generation focuses on designing novel structures by leveraging deep generative models trained on large-scale crystal datasets. These models, including Variational Autoencoders (VAEs), Generative Adversarial Networks (GANs), diffusion models, and Transformer-based architectures, have demonstrated significant potential in exploring vast chemical spaces and generating stable crystal structures. 
Notable examples include CDVAE~\cite{cdvae_xie2021crystal}, DiffCSP~\cite{diffcsp_jiao2023crystal}, CubicGAN~\cite{cubicgan_zhao2021high}, and MatterGPT~\cite{mattergpt_chen2024mattergpt}, each contributing to the advancement of generative materials design.
Building upon CDVAE, recent studies have incorporated multiple property constraints into the generation process, enabling property-oriented crystal design. Luo et al.~\cite{condcdvae_luo2024deep} introduced Cond-CDVAE, embedding composition and pressure conditions into the latent space to guide structure generation. Ye et al.~\cite{concdvae_ye2024cdvae} further refined this approach with Con-CDVAE, implementing a two-step training scheme that incorporates an additional network to model the latent distribution between crystal structures and multiple property constraints.
More recently, MatterGen~\cite{mattergen_zeni2025generative} has extended these capabilities by adopting a property-guided diffusion framework with classifier-free guidance~\cite{classifier_ho2022classifier}. Inspired by ControlNet~\cite{controlnet_zhang2023adding}, it integrates adapter modules for fine-tuning specific material properties, offering a more flexible and efficient approach to inverse crystal design. These advancements highlight the growing potential of deep generative models in accelerating the discovery of functional crystalline materials.

Despite these advancements, existing conditional crystal generative models face several limitations. Most notably, these models operate within a static generation paradigm, constrained by the distribution of their training datasets. As a result, their ability to explore novel chemical spaces beyond existing data remains limited. 
Furthermore, their performance often gets worse when trained on sparse datasets~\cite{concdvae_ye2024cdvae}, leading to inaccurate predictions in property regions that lack sufficient data coverage. This issue is particularly pronounced for materials with extreme property values or novel material classes that are underrepresented in current datasets~\cite{mattergpt_chen2024mattergpt, lu2022inverse, wang2025crystallinematerialdiscoveryera}.
To address these challenges, we propose an active learning framework that enhances the generative capabilities of conditional crystal models, particularly in sparsely-labelled data regions. Our framework employs a multi-layer screening process to dynamically refine and guide the generative model, ensuring more accurate and property-targeted crystal generation. 
As a case study, we apply this framework to improve the performance of Con-CDVAE~\cite{concdvae_ye2024cdvae} in generating crystal alloys with high bulk modulus (350 GPa). The effectiveness of our approach is validated using a predictive model based on CGCNN~\cite{cgcnn_xie2018crystal}, demonstrating the potential of active learning in expanding the reach of generative crystal design beyond conventional dataset limitations.

\section{Methods}
\label{sec:Methods}

\begin{figure*}[!htbp]
\includegraphics[width=0.95\textwidth]{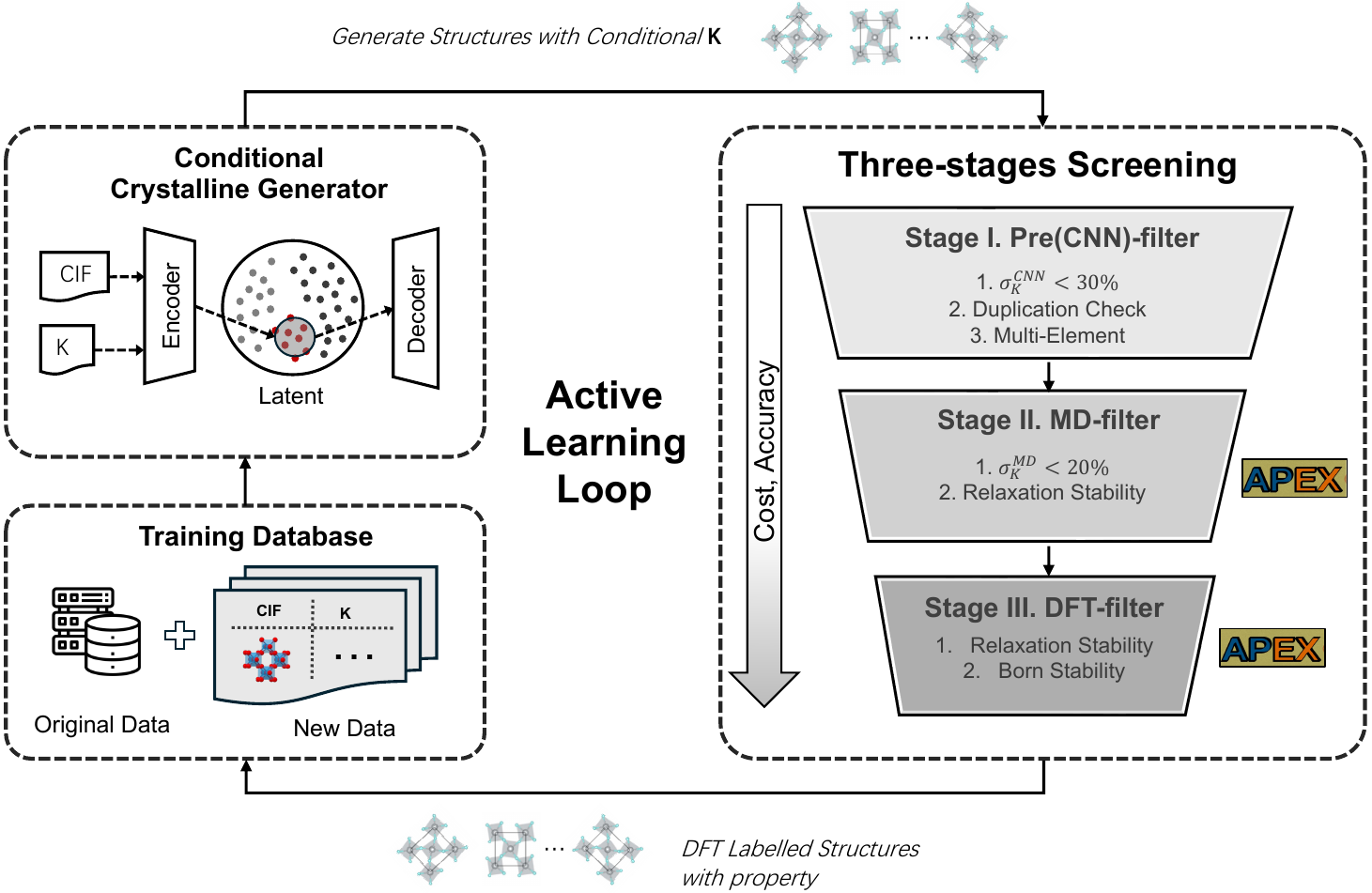}
\caption{\label{fig:loop} Active learning framework for inverse materials design. The active learning loop integrates a conditional crystal generation model with a three-stage screening process to iteratively refine crystal candidates. The screening stages include: (I) a graph neural network (GNN) for rapid property prediction, (II) molecular dynamics (MD) simulations using foundation atomic models, and (III) density functional theory (DFT) calculations for high-accuracy validation. MD and DFT simulations are performed using the Automatic Property Explorer (APEX)~\cite{apex_li2024extendable}. Validated structures and their computed properties are incorporated into the training dataset, progressively enhancing model accuracy and expanding the design space.}
\end{figure*}

\begin{figure*}[!htbp]
\includegraphics[width=0.95\textwidth]{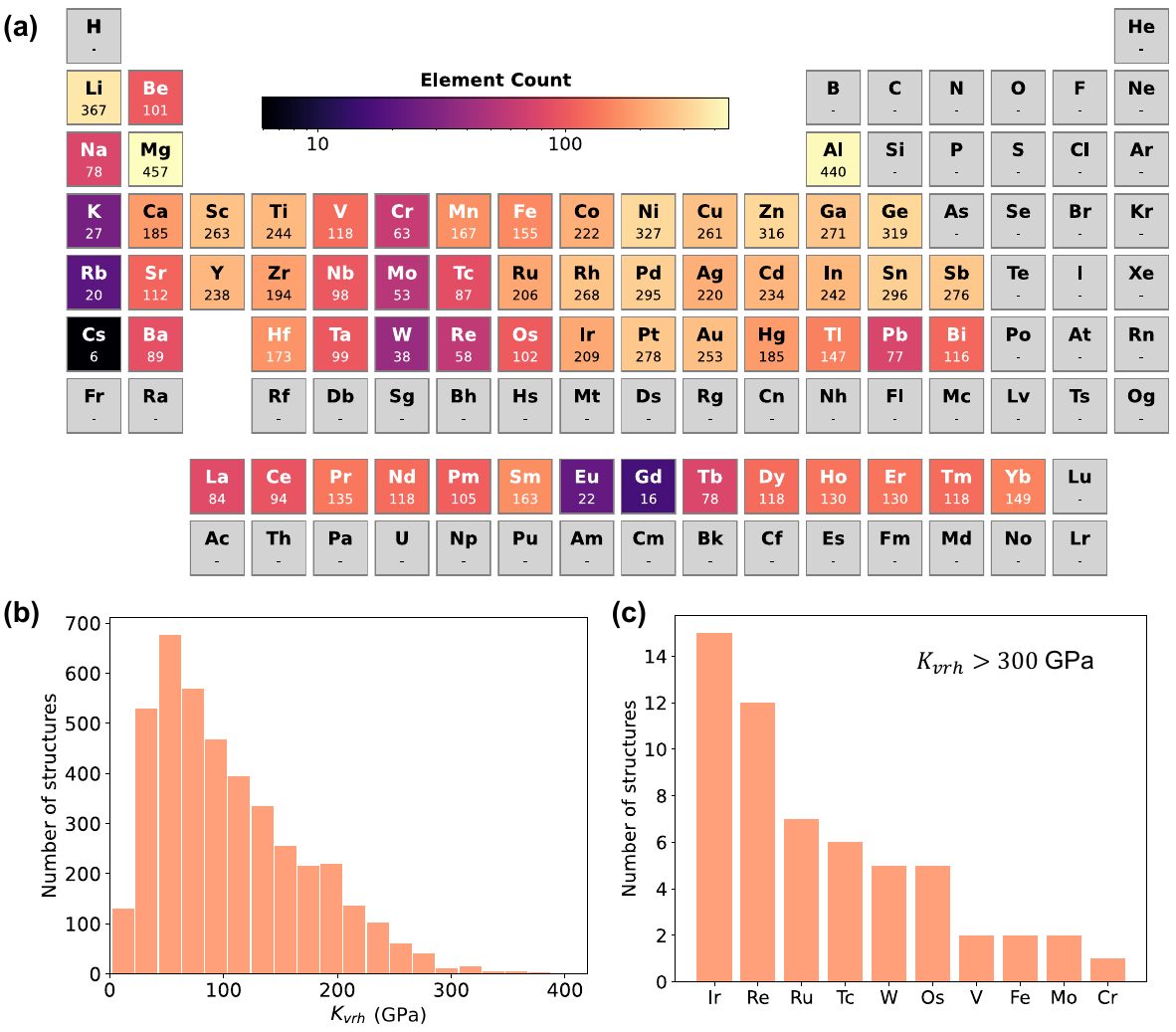}
\caption{\label{fig:data} Distribution of the initial training dataset. (a) Elemental occurrence frequency in the training dataset, illustrating the distribution of metal elements used for model training. (b) Histogram of bulk modulus $K_{\text {vrh}}$ values computed using the Voigt-Reuss-Hill (VRH) approximation, revealing a Poisson-like distribution with a peak between 40 and 80 GPa. (c) Distribution of crystal structures containing different elements with $K_{\text {vrh}}>$300 GPa, highlighting the prevalence of high-bulk-modulus materials, with Ir (Iridium) and Re (Rhenium) appearing most frequently.}
\end{figure*}

The active learning framework, illustrated in Fig.~\ref{fig:loop}, integrates a crystal structure generator with a three-stage screening process. This framework follows a semi-supervised learning approach, where newly generated crystal candidates undergo systematic filtering and property evaluation before being incorporated into the training dataset for iterative refinement. Specifically, the generative model proposes new structures, which are then assessed by the screening process, including first-principles calculations, to ensure accuracy and relevance. The validated data is subsequently added to the training set, enhancing the model’s learning in each iteration. Below, we describe each component in detail.

\subsection{Training Dataset}

The initial training dataset for bulk modulus prediction was derived from the MatBench v0.1 leaderboard (\verb|matbench_log_kvrh|)~\cite{matbench_dunn2020benchmarking, elas_data_de2015charting}, which compiles density functional theory (DFT)-calculated bulk modulus values from the Materials Project (MP)~\cite{elas_data_de2015charting}. To ensure data quality and reliability, we applied several preprocessing steps:
\begin{enumerate}
	\item Exclusion of unstable and nonphysical structures--Crystals with high formation energies ($>$150 meV), unrealistic modulus values (e.g., negative values or those exceeding 500 GPa due to inconsistencies in modulus approximation methods), and those containing noble gas elements were removed.
	\item Filtering by structural complexity--Crystals with more than 20 atoms per unit cell were excluded to reduce computational costs for subsequent DFT calculations.
	\item Focus on metallic alloys--Non-metallic elements and Actinides were filtered out to align the dataset with alloy-focused inverse design objectives.
\end{enumerate}

After applying these criteria, the refined dataset contained 5,296 structures spanning 62 metallic elements. The statistical distribution of element occurrences is presented in Fig.~\ref{fig:data}(a). The bulk modulus values, calculated using the Voigt-Reuss-Hill (VRH) approximation~\cite{vrh_hill_1952_ppssa}, follow a Poisson-like distribution, as shown in Fig.~\ref{fig:data}(b). The highest frequency of structures falls within the 40-80 GPa range, with over 600 entries, while the number of structures decreases progressively as the bulk modulus increases. Notably, structures with exceptionally high bulk modulus values ($>$300 GPa) are relatively rare, with fewer than 80 such entries, highlighting the data imbalance in high-stiffness materials.

Further analysis of the elemental distribution in high bulk modulus structures (Fig.~\ref{fig:data}(c)) reveals that Iridium (Ir) and Rhenium (Re) are the most prevalent elements, followed by Ruthenium (Ru), Technetium (Tc), Tungsten (W), and Osmium (Os). Vanadium (V), Iron (Fe), Molybdenum (Mo), and Chromium (Cr) appear with significantly lower frequency. For training the crystal generation model, the dataset was randomly split into training, validation, and test sets in an 8:1:1 ratio. This structured dataset serves as the foundation for iterative active learning cycles aimed at improving generative model accuracy for inverse materials design.

\subsection{Conditional Crystal Generator}
To generate crystal structures with targeted bulk modulus values, we adopt Con-CDVAE~\cite{concdvae_ye2024cdvae} as the conditional crystal generation model. Con-CDVAE employs a two-step training scheme, enabling property-constrained generation by embedding desired conditions into the latent space. Although Con-CDVAE supports multi-property training, we focus on bulk modulus $K_{\text {vrh}}$ as a single condition for this specific study.

In the first training phase, crystal structures and their corresponding $K_{\text {vrh}}$ values are encoded into a latent variable $z$, which is subsequently processed by a \textit{Decoder} for structure reconstruction and a \textit{Predictor} for property prediction.
The training objective is optimized by weighting the losses associated with different structural attributes: $p_{natom}=1$, $p_{coord}=10$, $p_{type}=1$, $p_{lat}=10$, $p_{comp}=1$, $p_{prop}=3$, where $p_{natom}$, $p_{coord}$, $p_{type}$, $p_{lat}$, $p_{comp}$, and $p_{prop}$ correspond to the number of atoms, atomic coordinates, elemental types, lattice parameters, composition, and target property, respectively. Training is performed for 600 steps on an Nvidia 4090 CPU. 

Following the initial training, we refine the latent space by training the \textit{Prior} blocks for 500 steps, using the same training dataset and property condition. This step enables the model to learn the distribution of latent variables corresponding to specific $K_{\text {vrh}}$ values, improving the generator’s ability to sample structures that align with the desired target properties.

Once the two-steps training is completed, the \textit{Prior} and \textit{Decoder} modules are employed to generate new alloy crystal structures based on specific bulk modulus values.
To ensure high-quality outputs, the \textit{Predictor} block is employed for post-generation filtering, selecting one candidate structure per 500 generated samples (500:1 filtering ratio). For each target bulk modulus input, we generate 500 candidate structures, with only the best-matching structures retained as the final output of Con-CDVAE.

\subsection{Three-Stage Screening Process}
To efficiently identify alloy crystals with the desired properties, we implement a three-stage screening framework that sequentially refines the generated candidates. The schematic workflow is illustrated in Fig.~\ref{fig:loop}, comprising the following filtering stages:

\begin{enumerate}
    \item{\textit{GNN pre-filter}}: The initially generated candidates from the conditional crystal generator undergo preliminary screening using a GNN model trained on the same dataset as Con-CDVAE. 
    This model predicts the bulk modulus $K_{\text {vrh}}$ directly from the crystal structure, and candidates with deviations exceeding 30\% from the target property are discarded. We employ CGCNN~\cite{cgcnn_xie2018crystal} for this task, ensuring efficient and reliable property screening.
    A duplication check using \verb|StructureMatcher| of Pymatgen~\cite{pymatgen_ong2013python} with default parameters (\verb|ltol|=0.2, \verb|stol|=0.3 and \verb|angle_tol|=5) ensures that redundant structures and single-element crystals are removed.

    \item{\textit{Molecular dynamics (MD)-filter}}: Candidates passing the GNN pre-filter are evaluated for stability and mechanical properties using MD simulations with the \verb|MACE-MP-0|~\cite{mace_batatia2023foundation} foundation atomic model in LAMMPS.
    High-throughput MD workflows are conducted via APEX~\cite{apex_li2024extendable}, where structures undergo full relaxation before elastic property calculations. Candidates exhibiting significant structural changes after relaxation are filtered out via \verb|StructureMatcher|, and only those with predicted bulk modulus values within 20\% of the target are retained.

    \item{\textit{DFT validation}}: The final selection of alloy candidates proceeds to DFT calculations for structural stability and mechanical validation. At this stage, both relaxation stability and elastic constants are assessed to ensure compliance with Born's stability criteria~\cite{born_PhysRevB.90.224104}. 
    The DFT joint (relaxation + elasticity) workflows are also conducted through APEX. The DFT convergence parameters are in accordance with the parameters to generate the training data~\cite{elas_data_de2015charting}.
    The DFT calculations are preformed using the Perdew-Burke-Ernzerhof~\cite{perdew_1996_prl} and generalized gradient approximation exchange-correlation functional with a plane-wave cutoff energy of 700 eV. 
    The projector-augmented-wave method~\cite{blochl_1994_prb} is employed to treat core and valence electrons. $K$-point sampling is implemented using the Monkhorst-Pack scheme~\cite{monkhorst_1976_prb}, with a grid spacing of 0.1 $\text{\AA}^{-1}$.  The convergence criteria for electronic minimization are set to be $10^{-5}$ meV between steps, while the residual force convergence criterion for ionic relaxation is set to 0.01 eV/\AA. Candidates meeting these criteria are added to the dataset for the next active learning iteration.
    
\end{enumerate}

This three-stage screening approach ensures that only structurally stable and property-optimized alloy candidates are selected for further refinement in the active learning cycle.

\section{Results and Discussion}
\label{sec:Results and Discussion}
\begin{figure*}[!htbp]
\includegraphics[width=0.95\textwidth]{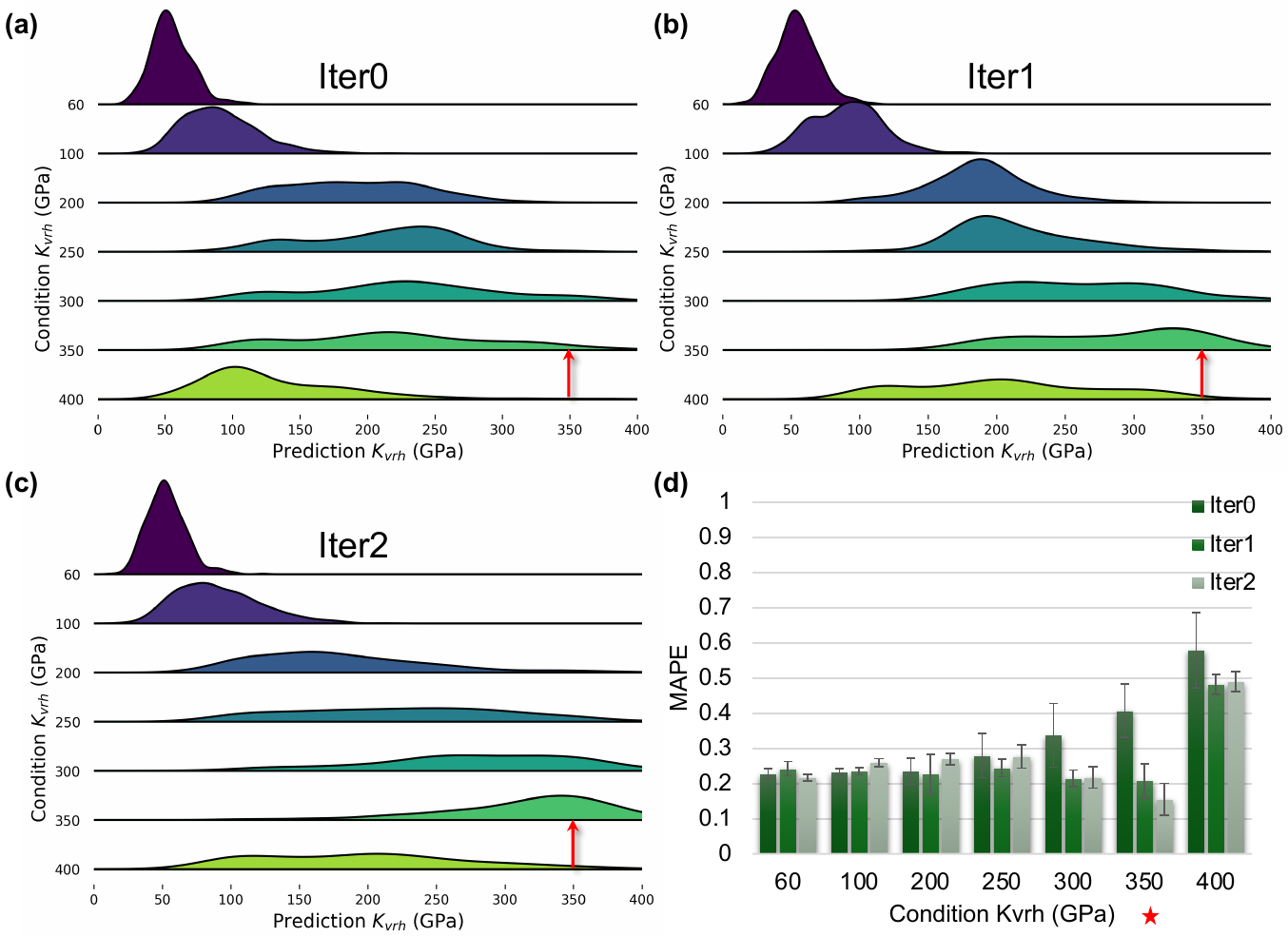}
\caption{\label{fig:distribution} Iterative improvement in bulk modulus prediction for generated alloy structures. (a–c) Bulk modulus predictions using CGCNN for generated structures across three iterations: (a) Iteration 0 (baseline performance), (b) Iteration 1 (first refinement), and (c) Iteration 2 (further optimization). Each horizontal line represents a target bulk modulus $K_{\text {vrh}}$, with corresponding CGCNN predictions shown along each line. (d) Evolution of the mean absolute percentage error (MAPE) across four models for different $K_{\text {vrh}}$ conditions, highlighting progressive accuracy improvements through the active learning process.}
\end{figure*}

To enhance the generation of alloy crystal structures with high bulk modulus $K_{\text {vrh}}$, we perform three iterations of active learning, targeting a bulk modulus condition of $K_{\text {vrh}}^{\text {target}}=$350 GPa, a region where data availability is limited in the original training dataset.

The generative model's performance is evaluated across multiple target conditions ($K_{\text {vrh}}$ = 60, 100, 200, 250, 300, 350, 300, 350, 400 GPa) by predicting the bulk modulus distribution for each iteration. 
This evaluation is conducted using the CGCNN model~\cite{cgcnn_xie2018crystal}, which is trained on the same dataset as the generative model (Con-CDVAE). 
The evolution of the generated distributions across iterations is shown in Figs.~\ref{fig:distribution}(a–c). To quantify model performance, we compute the mean absolute percentage error (MAPE) between the CGCNN-predicted values ($P_i$) and the target bulk modulus ($C$) (Fig.~\ref{fig:distribution}(d), using the following equation:

\begin{equation}\label{eq:mape}
	MAPE=\frac{1}{n}\sum_{i=1}^n\frac{|P_i-C|}{C}
\end{equation}
where $P_i$ is the predicted bulk modulus for the $i$-th sample, and $C$ is the target condition. 
To ensure statistical robustness, four independent models are trained for each condition, and the average MAPE is reported.
In the initial iteration (Iter0), the generative model is trained based on the original dataset, serving as a baseline for comparison (Fig.~\ref{fig:distribution}(a)).
The results indicate a strong correlation between the generation performance and the property distribution of the training data (Fig.~\ref{fig:data}(b)). 
When $K_{\text {vrh}}^{\text {target}}$ is 60 GPa, where the dataset is well-populated, the predicted values align closely with the target condition.
However, as $K_{\text {vrh}}$ increases, the generated distributions begin to deviate, particularly for conditions exceeding 250 GPa, where data sparsity becomes a limiting factor. At $K_{\text {vrh}}=$ 400 GPa, the model struggles to generate meaningful predictions, producing outputs that exhibit stochastic behavior without clear trends.

\begin{figure*}[!htbp]
\includegraphics[width=0.95\textwidth]{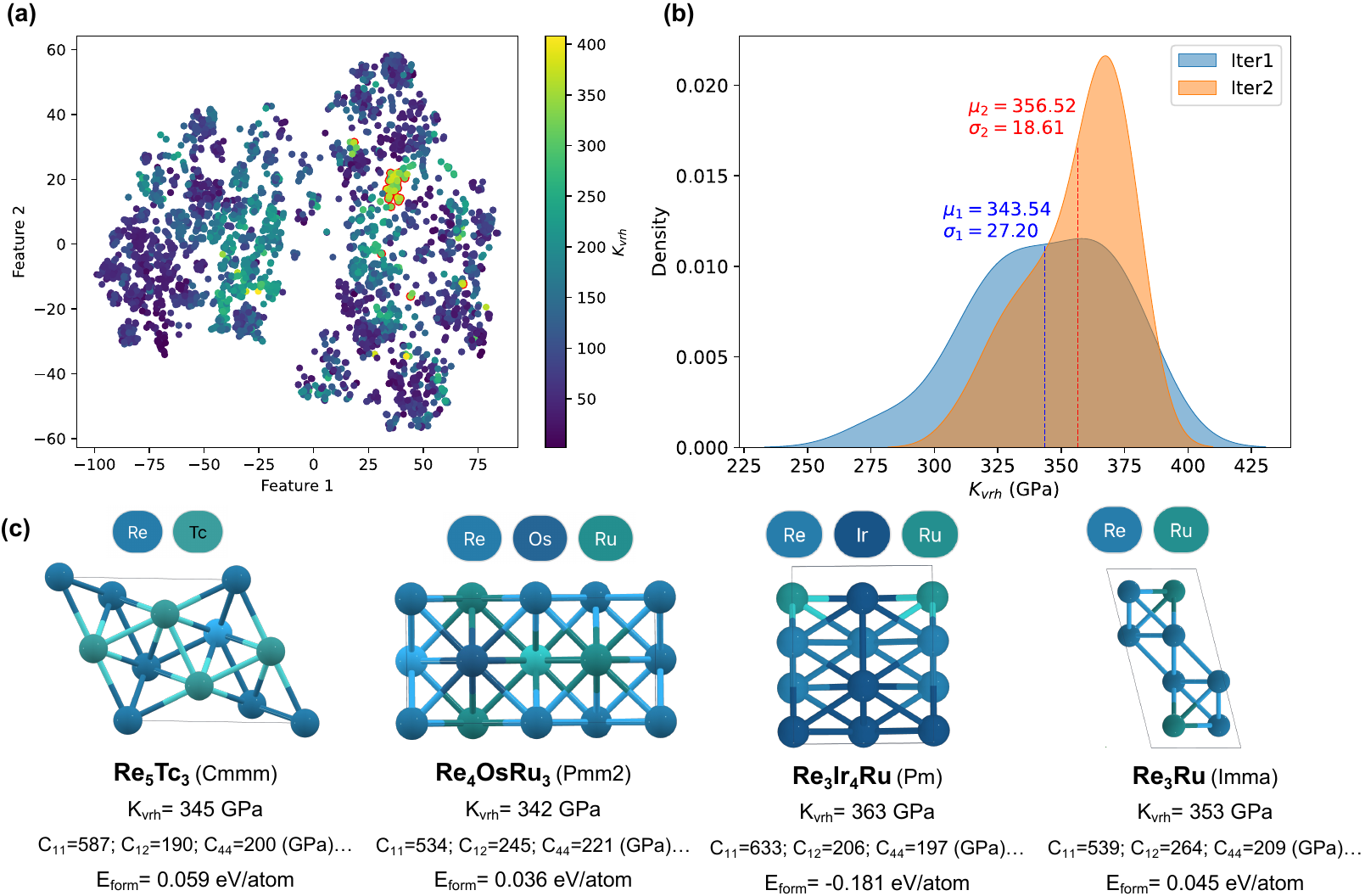}
\caption{\label{fig:dft} Newly identified alloy structures with high bulk modulus through active learning. (a) T-distributed Stochastic Neighbor Embedding (T-SNE) visualization of the latent space after two iterations, with newly generated high-bulk-modulus structures highlighted. (b) Distribution of DFT-calculated bulk modulus ($K_{\text {vrh}}$) for the newly generated structures across iterations, showing a shift toward the target value. (c) Four representative alloy crystals with DFT-validated bulk moduli near 350 GPa, demonstrating the effectiveness of the framework in discovering high-stiffness materials.}
\end{figure*}

Following the initial iteration, newly generated alloy crystals, labeled with their bulk moduli, are continuously incorporated into the training dataset to refine the generative model.
As shown in Figs.~\ref{fig:distribution}(a-c), the predicted $K_{\text {vrh}}$ distribution progressively clusters around the target value of 350 GPa (indicated by the red arrow). 
This trend is quantitatively reflected in the decrease of the average MAPE from 0.4 to 0.2, highlighting the effectiveness of the active learning loop in improving the accuracy of generate structures (Fig.~\ref{fig:distribution}(d)). 
Notably, the generative accuracy for conditions at 300 GPa and 400 GPa also improves, demonstrating the broader impact of the iterative refinement process. 
In the second iteration (Iter2), as additional alloy structures with bulk moduli near 350 GPa are incorporated into the training set, the MAPE further decreases to 0.14, indicating an increasingly refined predictive capability.
The consistency of the model’s predictions across other conditions--evaluated using CGCNN--suggests a stable generative performance, reinforcing the reliability of the framework.
This improvement is directly attributed to the iterative feedback mechanism, where each batch of newly generated structures refines and enhances the model’s predictive capacity, leading to progressively more accurate and property-aligned crystal structures.

To further analyze the structural evolution of the generated alloys, we visualize the latent space representation of the training dataset using T-distributed Stochastic Neighbor Embedding (T-SNE) after two iterations of active learning (Fig.~\ref{fig:dft}(a)). 
Newly added structures, highlighted by red circles, exhibit high bulk moduli ($K_{\text {vrh}}> 300$ GPa) and cluster into a newly explored region in the latent space. 
This clustering suggests that the framework is systematically expanding its search space toward high-modulus compositions, efficiently identifying unexplored regions with promising material candidates.

The distribution of DFT-calculated $K_{\text {vrh}}$ for newly generated structures is shown in Fig.~\ref{fig:dft}(b). 
After the first iteration (Iter1), the mean bulk modulus of generated structures is $\mu_0=343.54$ GPa with a standard deviation of 27.20 GPa, indicating a relatively broad spread around the target value. 
The slight underestimation of the target modulus suggests a conservative initial generation process. However, after a subsequent iteration of active learning (Iter2), the mean $K_{\text {vrh}}$ shifts to 356.52 GPa, accompanied by a significant reduction in standard deviation to 18.61 GPa, implying enhanced consistency and alignment with the target property.

Fig.~\ref{fig:dft}(c) shows four examples of newly generated alloy crystals with DFT-labeled bulk moduli near 350 GPa. Most of these structures are rhenium (Re)-based, combined with other high-atomic-stiffness elements such as technetium (Tc), osmium (Os), ruthenium (Ru), and iridium (Ir) to form binary and ternary alloy compounds. These findings align well with prior empirical studies on ultra-incompressible crystals~\cite{jin2023atomic} and the elemental distribution in the initial training dataset (Fig.~\ref{fig:data}(c)). Notably, these newly identified crystals are absent from the Materials Project database~\cite{elas_data_de2015charting}, yet they satisfy Born’s mechanical stability criteria and exhibit low/negative formation energies for thermodynamic stability.

These results demonstrate that the active learning-driven inverse design framework effectively refines crystal generation, allowing for targeted exploration of sparsely populated regions in property space. By iteratively guiding the generative model with high-quality training data, the framework significantly enhances both the accuracy and efficiency of discovering novel crystalline materials with tailored mechanical properties.

\section{Conclusion}
In this study, we present an active learning framework for inverse materials design, integrating crystal generation models and foundation atomic models to efficiently identify new crystalline materials with targeted properties.
Traditional approaches for exploring the vast compositional space of crystalline materials are often constrained by high computational costs and time-intensive experimental validation. To address these limitations, our framework employs a three-stage screening mechanism that dynamically refines the generative process, leveraging AI-driven techniques to accelerate materials discovery.

As a case study, we implement Con-CDVAE, a conditional crystal generation model, and fine-tune it to design alloy structures with a high bulk modulus of 350 GPa. While we demonstrate our approach using Con-CDVAE, the framework is model-agnostic and can seamlessly incorporate other generative models, such as the recently developed MatterGen~\cite{mattergen_zeni2025generative}. Our active learning loop iteratively refines predictions and enriches the training dataset, enhancing both the accuracy and efficiency of the generation process. 
The generated candidates undergo three layers of screening: (1) a graph neural network (GNN)-based pre-filter, (2) stability checks via MD simulations using foundation atomic models, and (3) final validation through DFT calculations. With the continued advancement of foundation atomic models, the overall performance of this framework is expected to improve further, making it increasingly efficient for inverse materials design.

Beyond its current implementation, the active learning framework introduced here serves as a versatile platform that integrates two key AI-driven methodologies in materials science--crystal generation and foundation atomic models. The framework is designed for adaptability, allowing seamless integration with emerging AI techniques, including large language models (LLMs) and AI agents~\cite{liu2024prompt,hu2024multi}. By incorporating literature mining and autonomous decision-making and reasoning capabilities, future iterations of this framework could establish a fully automated materials design pipeline.

Overall, our results demonstrate that active learning with advanced AI materials models can significantly enhance both the accuracy and speed of materials discovery, addressing data sparsity issues and optimizing inverse design processes. This study highlights the transformative role of AI in accelerating materials research, paving the way for data-driven, autonomous materials design workflows that push the boundaries of materials science.

\section{DATA AVAILABILITY}
The data and code used in this work are available upon reasonable request. 

\section{ACKNOWLEDGMENTS}

This work is supported by Research Grants Council, Hong Kong SAR through the General Research Fund (17210723, 17200424). T.W. acknowledges additional support by The University of Hong Kong (HKU) via seed fund
(2409100597).

\def\bibsection{\section*{\refname}} 

\bibliography{./main.bib}

\newpage

\end{document}